\documentstyle[general_cite,psfig]{mn}
\title{The intragroup medium in loose groups of galaxies}
\author[Stephen F. Helsdon and Trevor J. Ponman]
       {Stephen F. Helsdon\thanks{E-mail: sfh@star.sr.bham.ac.uk} and Trevor J.
  Ponman \\
  School of Physics and Astronomy, University of
        Birmingham, Edgbaston, Birmingham B15 2TT, UK\\}
 \date{Accepted 1999 ??.
      Received 1999 ??;
      in original form 1999 ??}

\pagerange{\pageref{firstpage}--\pageref{lastpage}}
\pubyear{1998}

\def\deg{\hbox{$^\circ$}}
\def\spose#1{\hbox to 0pt{#1\hss}}
\def\gtsim{$\mathrel{\spose{\lower 3pt\hbox{$\sim$}}
        \raise 2.0pt\hbox{$>$}}$} 
\def\ltsim{$\mathrel{\spose{\lower 3pt\hbox{$\sim$}}
        \raise 2.0pt\hbox{$<$}}$}

\begin{document}

\maketitle

\label{firstpage}

\begin{abstract}

  \noindent We have used the {\it ROSAT} PSPC to study the properties of a
  sample of 24 X-ray bright galaxy groups, representing the largest sample
  examined in detail to date. Hot plasma models are fitted to the spectral
  data to derive temperatures, and modified King models are used to
  characterise the surface brightness profiles.
  
  In agreement with previous work, we find evidence for the presence of two
  components in the surface brightness profiles.  The extended component is
  generally found to be much flatter than that observed in galaxy clusters,
  and there is evidence that the profiles follow a trend with system mass.
  We derive relationships between X-ray luminosity, temperature and optical
  velocity dispersion. The relation between X-ray luminosity and
  temperature is found to be $L_X \propto T^{4.9}$, which is significantly
  steeper than the same relation in galaxy clusters. These results are in
  good agreement with preheating models, in which galaxy winds raise the
  internal energy of the gas, inhibiting its collapse into the shallow
  potential wells of poor systems.

\end{abstract}


\section{Introduction}
\label{sec:intro}

The majority of galaxies in the universe are found in galaxy groups
\cite{tully87}. These collections of between 3 and about 30 galaxies trace
large scale structure (Ramella, Geller \& Huchra 1990)\nocite{ramella90}
and probably contain a large fraction of the total baryonic mass in the
universe (Fukugita, Hogan \& Peebles 1998)\nocite{fukugita98}.  However
despite their abundance and importance, galaxy groups have received
relatively little attention until recently. The main problem has been the
identification of the groups themselves. Even when redshift information is
available, it is difficult to identify whether a group is truly bound, due
to the problems of small number statistics and chance superpositions. In
contrast, galaxy clusters which are easier to identify due to the larger
number of members, have been extensively studied.

The detection of extended X-ray emission from hot gas in the group
potential well provides the best evidence that a group is truly
gravitationally bound. The study of this hot intragroup gas can provide
important insights into the evolution and dynamics of the group and its
member galaxies. Samples of X-ray bright groups were originally studied
using the {\it Einstein} satellite (e.g. \pcite{price91}), but the
introduction of the {\it ROSAT} satellite with its improved sensitivity and
resolution, allowed a more thorough analysis of these systems. Since the
{\it ROSAT} PSPC was first used to study X-ray bright groups
\cite{mulchaey93,ponman93a} a number of collections of groups have been
studied (e.g.
\pcite{doe95,ponman96a,burns96,mulchaey96,mahdavi97,mulchaey98b}).
However, none of these studies provides a uniform, detailed analysis of a
reasonable sized sample of groups, based on high quality data. The largest
samples have all been based on {\it ROSAT} All Sky Survey (RASS) data, in
which case properties other than the luminosities are difficult to
determine, due to poor statistics resulting from the short exposures.

The study of \scite{ponman96a} used a mixture of RASS and pointed data and
identified 22 X-ray bright groups. These were all compact groups from the
catalogue of \scite{hickson82}. Such compact groups have the advantage that
they can be easily identified on the sky due to the high projected
over-densities of galaxies within them, but may be unrepresentative of
groups as a whole.  The X-ray properties of these Hickson compact groups
(HCGs) showed systematic departures from those of clusters, leading to the
suggestion that they might be displaying the marks of energy injection into
the intergalactic medium due to galaxy winds.

\scite{mulchaey98b} (henceforth MZ98) used pointed PSPC data to study
groups of both loose and compact morphology.  With a sample of only nine
groups they were unable to derive reliable statistical results, however
they found that properties such as the $L:T$ relation and surface
brightness slope were indistinguishable from those of clusters, in
contradiction to the results of \scite{ponman96a}. If this is true, it
suggests a fundamental difference between the properties of loose and
compact groups. The main aim of the present work is to establish the X-ray
properties of loose groups by means of a careful and uniform study of a
larger sample of systems, and to establish whether they do, in fact, differ
from compact groups in their X-ray properties. To allow direct comparison
with the results of MZ98, their sample has been included within ours.

In \S~\ref{sec:sample_selection} we describe the sample selection and
initial identification of the X-ray bright groups. The spectral and spatial
analyses of the X-ray emission are described in
\S~\ref{sec:spectral_analysis} and \S~\ref{sec:spatial_analysis}.  Results
of the analysis, including correlations between the derived parameters, are
presented in \S~\ref{sec:Distributionsandcorrelations}.  These results are
compared with those of MZ98 in \S~\ref{Comparison}, and discussed in
\S~\ref{Discussion}. Finally, our conclusions are summarized in
\S~\ref{Conclusions}. Throughout this paper we use H$_0$~=~50~km~s$^{-1}$
Mpc$^{-1}$.

\section{Sample Selection and Data Reduction}
\label{sec:sample_selection}

The primary aim of this work is to study the properties of a number of
X-ray bright groups, as such it was necessary to initially compile such a
sample. Three different sources were used for this purpose, the optical
catalogue of \scite{nolthenius93}, the sample of \scite{ledlow96} and the
X-ray bright groups from MZ98. The catalogue of \scite{nolthenius93}
contains 173 groups, with three or more members, selected from the CfA1
galaxy redshift catalogue using a friends of friends algorithm with a
density enhancement of 15. The \scite{ledlow96} sample contains 71 groups
selected from the Zwicky Catalogue of Galaxies and Clusters of Galaxies,
using a friends of friends algorithm with a surface density enhancement of
46.4, each group having at least 4 members and galactic latitude $|b|~\geq
30\deg$.

Cross-correlation of the \scite{nolthenius93} and \scite{ledlow96} samples
with the {\it ROSAT} observing log, identified groups which had been
observed by the {\it ROSAT} PSPC during its programme of pointed
observations.  We further restricted ourselves to groups which had been
observed within 20$'$ of the centre of the PSPC. The nine X-ray bright
groups from MZ98 were all known to have been observed by the {\it ROSAT}
PSPC and were added to the sample.  Groups identified as being part of
known bright galaxy clusters such as Coma were also excluded at this stage.
This resulted in a potential sample of 37 galaxy groups, which are listed
in Table~\ref{tab:allgroups}.

Before the X-ray data can be used it is necessary to identify and exclude
sources of contamination. These include particle events and solar X-ray
emission scattered from the Earth's atmosphere into the telescope.
Detectors on board the spacecraft identify and exclude over 99\% of the
particle events that would be recorded as X-rays. These particle events are
recorded as the master veto rate. At values of the master veto rate of
above 170 count s$^{-1}$ the contamination by particles is significant, and
these times are excluded from our analysis. Reflected solar X-rays can be
identified by an increase in the total X-ray event rate. To remove this
contamination, times where the total event rate deviated by more than
2$\sigma$ from the mean were excluded. Typically this resulted in the
removal of a few percent of each observation.

A standard reduction of the data was then carried out to produce an image
and background for each group. The statistical significance of any emission
within distances of 50~kpc and 200~kpc from the optical centre of each of
the groups was then calculated. This was used along with a smoothed image
and a profile of the group to identify the presence of extended emission
above a 5$\sigma$ detection threshold. It was also apparent that in a few
cases diffuse X-ray emission was centred on a galaxy within the PSPC ring,
even though the catalogued optical centre of the group was outside the
ring. These groups were also included, and are identified with an asterisk
in Table~\ref{tab:allgroups}. This resulted in a final sample of 24 X-ray
bright galaxy groups, which are identified in Table~\ref{tab:allgroups}. As
the table shows, those systems span a considerable range in catalogued
optical richness ($N_{gal} = 3-45$). A more reliable measure of the total
mass of each group is given by the X-ray temperatures derived below. Our
sample should not be regarded as being statistically complete in any way,
but rather a reasonably representative sample of X-ray bright groups.

\begin{table*}
\begin{minipage}[c]{17cm}
\caption{\label{tab:allgroups}Listed are the groups in which a
  search for extended X-ray emission was carried out. Groups with a 1 in
  the comments column have properties taken from Nolthenius (1993), those
  identified with a 2 are from Ledlow et al. (1996) and those marked with a
  3 are from Mulchaey \& Zabludoff (1998). Asterisks indicate groups in
  which emission was identified within the PSPC support ring, but whose
  catalogued optical positions were outside the ring. Groups with detected
  X-ray emission are listed in the top half of the table along with the
  radius to which emission was observed. Groups that were not used are
  given in the lower region of the table along with the reason for
  exclusion.}  \center{
\begin{tabular}{llccrrclc}
\hline
Name     & Alt. Name         & RA(2000)     &   Dec(2000)     & $N_{gal}$ & $\sigma$ (km s$^{-1}$) & z & Comments & R$_{ext}$ ($'$)                  \\
\hline                                                      
NGC 315  & Nol 6             & 00 58 25.0   &   +30 39 11     & 4  & 122 & 0.0164 & 1 $\ast$ & 6.0\\
NGC 383  & S34-111           & 01 07 27.7   &   +32 23 59     & 29 & 466 & 0.0173 & 2  & 30.0   \\
NGC 524  & Nol 11            & 01 24 01.6   &   +09 27 37.7   & 8  & 205 & 0.0083 & 1  & 10.6   \\
NGC 533  &                   & 01 25 29.1   &   +01 48 17     & 36 & 464 & 0.0181 & 3  & 20.3   \\
NGC 741  & S49-140           & 01 57 00.7   &   +05 40 00     & 41 & 432 & 0.0179 & 3  & 16.0   \\
NGC 1587 & Nol 33            & 04 30 46.1   &   +00 24 25.7   & 4  & 106 & 0.0122 & 1  & 6.0   \\
NGC 2563 & NGC 2563          & 08 20 24.4   &   +21 05 46     & 29 & 336 & 0.0163 & 3  & 17.6   \\
NGC 3091 & HCG 42            & 10 00 13.1   &   -19 38 24     & 22 & 211 & 0.0128 & 3  & 8.9   \\
NGC 3607 & Nol 65            & 11 17 55.9   &   +18 07 35.8   & 3  & 421 & 0.0037 & 1  & 9.6   \\
NGC 3665 & Nol 68            & 11 23 30.6   &   +38 43 31.6   & 4  & 29  & 0.0069 & 1  & 6.0   \\
NGC 4065 & N79-299A,Nol 91   & 12 04 09.5   &   +20 13 18     & 9  & 495 & 0.0235 & 2  & 15.0   \\
NGC 4073 & N67-335           & 12 04 21.7   &   +01 50 19     & 22 & 607 & 0.0204 & 2  & 18.0   \\
NGC 4261 & Nol 99,N67-330    & 12 20 02.3   &   +05 20 24     & 33 & 465 & 0.0071 & 1  & 15.0  \\
NGC 4325 & NGC 4325          & 12 23 18.2   &   +10 37 19     & 18 & 256 & 0.0252 & 3  & 10.2   \\
NGC 4636 & Nol 104           & 12 42 57.2   &   +02 31 34.3   & 12 & 463 & 0.0044 & 1  & 21.6   \\
NGC 4761 & HGC62             & 12 52 57.9   &   -09 09 26     & 45 & 376 & 0.0146 & 3  & 15.6   \\
NGC 5129 & Nol 117           & 13 24 36.0   &   +13 55 40     & 33 & 294 & 0.0232 & 3  & 9.0   \\
NGC 5171 & N79-296           & 13 29 22.3   &   +11 47 31     & 8  & 424 & 0.0232 & 2  & 10.8   \\
NGC 5353 & Nol 124,N79-286,HCG68  & 13 51 37.0   &   +40 32 12& 15 & 174 & 0.0081 & 1 $\ast$  & 9.6\\
NGC 5846 & Nol 146           & 15 05 47.0   &   +01 34 25     & 20 & 368 & 0.0063 & 3  & 15.0   \\
NGC 6338 & N34-175           & 17 15 21.4   &   +57 22 43     & 7  & 589 & 0.0283 & 2  & 13.8   \\
NGC 7176 & HCG90             & 22 02 31.4   &   -32 04 58     & 16 & 193 & 0.0085 & 3  & 13.5   \\
NGC 7619 & Nol 164           & 23 20 32.1   &   +08 22 26.5   & 7  & 253 & 0.0111 & 1  & 24.0   \\
NGC 7777 & Nol 170$^{\ast}$  & 23 53 33.0   &   +28 34 42     & 4  & 116 & 0.0229 & 1 $\ast$ & 6.6 \\
\hline
NGC 7819 & Nol 173               & 00 02 28.0   &   +31 28 42.1   & 3  & 71  & 0.0164 & \multicolumn{2}{l}{1  no detection}    \\
NGC 43   & Nol 1                 & 00 13 05.8   &   +30 58 40.8   & 3  & 63  & 0.0160 & \multicolumn{2}{l}{1  no detection}     \\
NGC 2769 & Nol 35                & 09 10 22.8   &   +50 23 45.3   & 3  & 125 & 0.0166 & \multicolumn{2}{l}{1  no detection}    \\
NGC 3839 & Nol 82,N67-312        & 11 42 04.6   &   +10 18 20.0   & 9  & 177 & 0.0206 & \multicolumn{2}{l}{2  background clusters} \\
NGC 4168 & Nol 98                & 12 13 38.9   &   +13 01 19.3   & 4  & 152 & 0.0077 & \multicolumn{2}{l}{1  no detection}    \\
NGC 4360 & Nol 101               & 12 25 44.6   &   +09 07 23.5   & 3  & 289 & 0.0245 & \multicolumn{2}{l}{1  behind Virgo emission} \\
NGC 4615 & Nol 108               & 12 41 16.0   &   +26 13 33.2   & 3  & 47  & 0.0158 & \multicolumn{2}{l}{1  too few counts}  \\
NGC 5386 & Nol 129               & 13 58 00.1   &   +06 15 25.1   & 3  & 9   & 0.0143 & \multicolumn{2}{l}{1  no detection}    \\
NGC 5775 & Nol 143               & 14 53 24.9   &   +03 29 47.5   & 5  & 88  & 0.0051 & \multicolumn{2}{l}{1  no detection}    \\
NGC 5866 & Nol 147               & 15 16 23.5   &   +56 25 01.9   & 4  & 74  & 0.0022 & \multicolumn{2}{l}{1  no detection}    \\
NGC 5970 & Nol 154               & 15 36 16.2   &   +12 02 07.7   & 3  & 81  & 0.0064 & \multicolumn{2}{l}{1  no detection}    \\
NGC 7448 & Nol 160,S49-143       & 23 01 48.9   &   +15 58 09.0   & 8  & 153 & 0.0077 & \multicolumn{2}{l}{2  no detection}    \\
\hline
\end{tabular}
}
\end{minipage}
\end{table*}

\nocite{nolthenius93,ledlow96,mulchaey98b}

\section{Spectral analysis}
\label{sec:spectral_analysis}

Events surviving the initial screening process were binned into a
3-dimensional $x,y,Energy$ data cube. An estimate of the background was
generated from an annulus at r=0.6-0.7\deg with the PSPC support spokes
removed. The dataset was then background subtracted, and point sources
identified using a maximum likelihood source searching program.  Point
sources within the background annulus were removed to 1.2 times the 95\%
radius for 0.5~keV photons. The background was then recalculated and the
image once again searched for point sources. Other more extended sources,
such as background galaxy clusters not associated with the group emission,
were also manually identified and excluded at this point. This process of
identifying and removing point sources to produce a better estimate of the
background was repeated until the same number of point sources was
identified each time. Typically this took 4-5 iterations for each dataset.

The final background subtracted data were then corrected for dead time
effects and vignetting, and then divided by the effective exposure time to
give a map of spectral flux. A circular region around each of the groups
was used to extract a spectrum. The size of this region was determined by
examination of a smoothed image and a surface brightness profile of the
group. The region was selected to include all the emission that could be
observed in the smoothed image and profile; its size for each of the groups
is shown in Table~\ref{tab:allgroups}.  Point sources, and other sources as
identified above, were removed from the spectral image, along with the
support structure and the data outside the radius of interest. The spectrum
for each group was then obtained by collapsing the spectral image along the
$x$ and $y$ axes.

Each spectrum was fitted with a MEKAL hot plasma model (Mewe, Lemen \& van
den Oord 1986)\nocite{mewe86a} with a hydrogen absorbing column frozen at a
value determined from radio surveys \cite{stark92a}. For two of the groups
it was also necessary to fix the abundance to obtain a sensible fit. A
value of 0.3 solar was used for this purpose. In this way we derived
temperature, abundance and bolometric flux for each group.

For hot spectra, the limited spectral band of {\it ROSAT} makes temperature
determination subject to systematic errors in the high energy response of
the PSPC, and there is evidence that {\it ROSAT} temperatures are
systematically lower than those from hard X-ray instruments such as {\it
Ginga} and {\it ASCA}.  A comparison of {\it ROSAT} and {\it ASCA}
temperatures by \pcite{hwang99} showed that this temperature bias amounts
to $\sim 30$\% in hot systems, but that there is no evidence of any
systematic offset below an {\it ASCA} derived temperature of 2keV, where
the {\it ROSAT} band covers the spectrum adequately.  We therefore expect
the {\it ROSAT}-determined temperatures for the systems in our sample
(which have $T<1.7$~keV) to be free from serious bias.

The distribution of group temperatures in the present sample occupies a
rather small range around 1~keV. For the 22 groups in which metallicities
were derived, the overall weighted mean metallicity is $0.19 \pm 0.01$
solar, whilst the median is 0.42 solar. A trend is observed in clusters for
higher metallicity in lower temperature systems \cite{arnaud94}. This would
lead one to expect a typical metallicity of $\approx$ 0.6 solar in systems
with $T \approx 1$ keV. However there is evidence that there may be
abundance gradients in cool clusters which result in an increased abundance
at the centre (e.g.  \pcite{xu97,ikebe97,fukazawa98,finoguenov99}) and
could account for the observed trend. In any case, results obtained here
for the group metallicities must be viewed with caution, since {\it ROSAT}
is unable to resolve individual emission lines, and metallicities can be
strongly biased when a variable temperature plasma is fitted with an
isothermal model (e.g. Buote \& Fabian 1998\nocite{buote98}, Finoguenov \&
Ponman 1999).

\begin{figure*}
\hspace{0cm}
\psfig{file=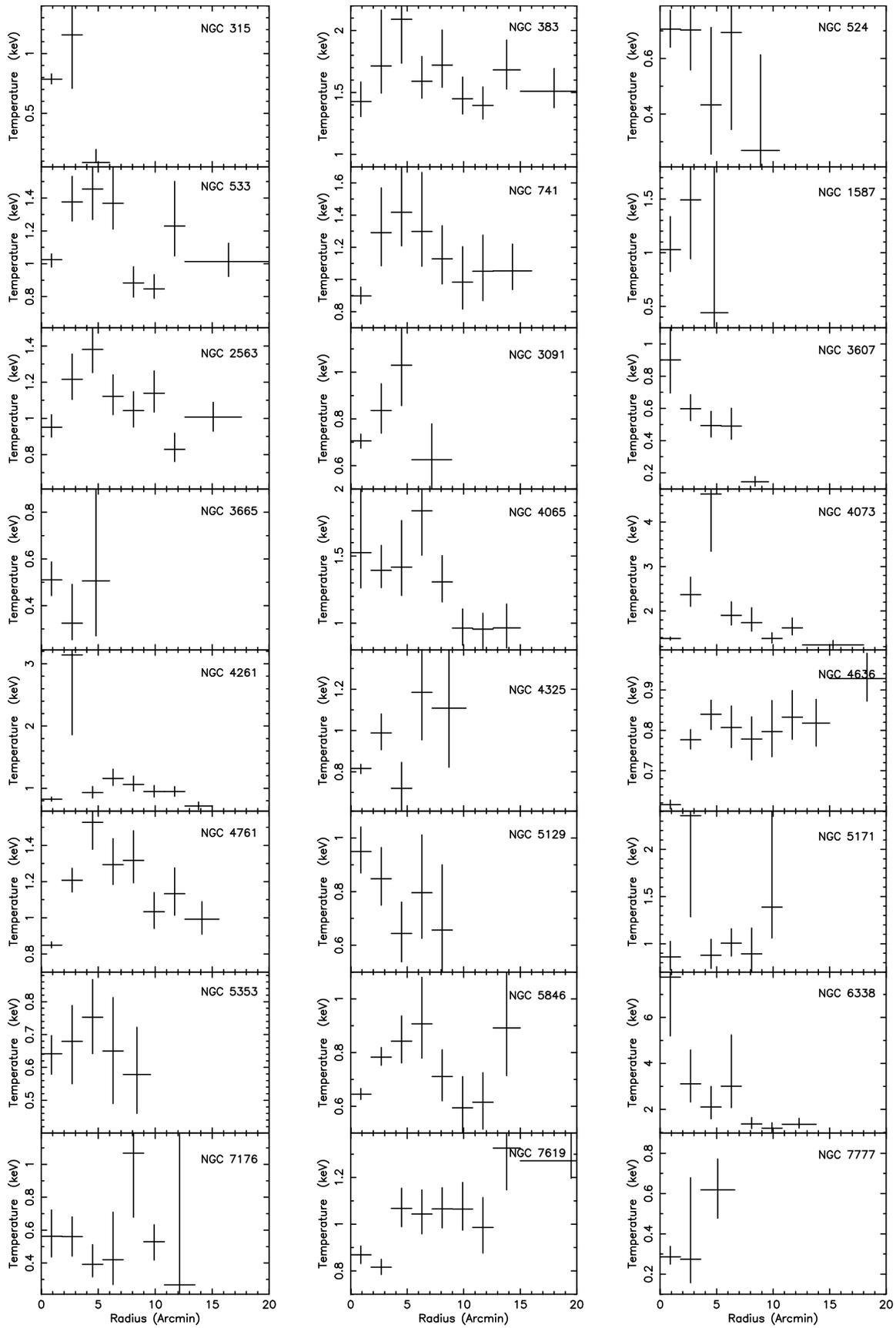,angle=0,width=16cm}
\caption{\label{fig:profiles}Temperature profiles from an annular analysis
  of the 24 groups in this sample. A number of the groups show a
  temperature drop in the central regions and/or at large radii.}
\end{figure*}

For each of the groups in the sample we also derived simple projected
temperature profiles. In each case spectra in several annuli were extracted
and fitted with a MEKAL model as described above. In each annulus the
hydrogen column and abundance were frozen at the global values. The
resulting temperature profiles of all groups are shown in
Fig.\ref{fig:profiles}. Some of the profiles shown are not particularly
informative due to a combination of large errors on the temperatures and a
small number of annuli. However it is clear that approximately half of the
groups show evidence of a temperature drop in the central regions,
indicating the presence of a cooler component. Also, approximately half of
the profiles show evidence of a decline in temperature at large radii.

\section{Surface Brightness profiles and Group Luminosities}
\label{sec:spatial_analysis}

Observations of galaxy clusters across a wide range in virial temperature
appear to indicate a flattening of the profiles in lower mass systems
(\pcite{arnaud99}; Ponman, Cannon \& Navarro 1999\nocite{ponman99}) -- a
result consistent with expectations if the intergalactic gas has been
subject to preheating by galaxy winds \cite{metzler99,cavaliere99}.
However, MZ98 found for their sample of groups that surface brightness
profiles did not differ significantly from those in clusters, once the
presence of central components was properly allowed for.  We set out to
examine the surface brightness profiles of our group sample in an attempt
to resolve this issue.

Following initial reduction, an image was extracted in the 0.5 - 2 keV
band, and corrected for vignetting using an energy dependent exposure map
(see \scite{snowden94a} for description). Point sources identified in the
spectral analysis were removed from the image along with any other
unrelated extended sources. Only the data within the region from which each
group spectrum was extracted, were used for the spatial analysis. It has
been shown that the centroid of the X-ray emission often lies at the
position of the brightest group galaxy (MZ98), and as such any emission
centred on this galaxy may be associated with the group potential as a
whole. For this reason any source associated with the centre of the X-ray
emission was not removed. Use of the energy dependent exposure map to
correct for vignetting, results in a constant background level across the
image, therefore a flat background was also determined and subtracted from
the data.

For each group the 2-dimensional surface brightness profile was modelled
with a modified King function (or '$\beta$-profile')of the form:\\

\noindent
\begin{math}
S(r)=S_0(1+(r/r_{core})^{2})^{-3\beta_{fit}+0.5}
\end{math}
\\

Models were convolved with the PSPC point spread function at an energy
determined from the mean photon energy of the group spectrum, and fitted to
the data. The free parameters were the central surface brightness $S_0$,
the core radius $r_{core}$, the index $\beta_{fit}$ and the $x$ and $y$
position of the centre of the emission. Both spherical and elliptical fits
were carried out on the data, with the major to minor axis ratio and the
position angle being extra free parameters in the elliptical fits.

The use of 2-dimensional datasets to fit the surface brightness
distribution results in a low number of counts in many of the data bins.
Under these conditions chi-squared ($\chi^2$) fitting performs poorly, so
maximum likelihood fitting, using the Cash statistic, was used instead. The
Cash statistic \cite{cash79} is defined as $-2{\rm ln}L$ where $L$ is the
likelihood function (in this case derived from the Poisson distribution).
Thus the most likely model has a minimum Cash statistic. Differences in the
Cash statistic are $\chi^2$ distributed, so confidence intervals may be
calculated in the same way as for a conventional $\chi^2$ fit.

Unfortunately the Cash statistic by itself gives no indication of the
quality of a fit; hence it was necessary to obtain some other estimate of
the fit quality. A Monte Carlo approach was used, in which the best fit
model was used to simulate 1000 images of the group. Poisson noise was
added to each of these images, and they were then compared to the original
model, and the Cash statistic for each image determined.  Thus, for a
particular model we were able to obtain a distribution showing the range of
Cash values expected for datasets generated from this model. A Gaussian was
then fitted to this distribution to obtain the width and central value. By
comparing the Cash statistic for the real dataset with this distribution,
it was possible to determine the probability that the model could have
produced the data. This probability is recorded in
Table~\ref{tab:2kingfit}, as the number of standard deviations that the
real value lies from the centre point of the distribution.  If the value of
the real Cash statistic lay more than 2$\sigma$ from the peak of the
distribution then the fit was regarded as `poor'.

As can be seen in Table~\ref{tab:2kingfit}, the single-component fits
provide an adequate description of the data in a few cases. However for
most groups the single-component fits are poor. It has been suggested that
there are typically two components in the surface brightness profiles of
galaxy groups (MZ98), a central component associated with a central galaxy,
cooling flow or AGN, and a more extended component associated with the
group potential. To check this, models comprising of two superposed
$\beta$-profiles were also fitted to those datasets with poor
single-component fits and greater than $\approx900$ total counts. Below
this number of source counts, statistics were found to be too poor to
constrain the more complicated two-component models. To limit the number of
free parameters, the central component was constrained to be spherical
while the outer component was allowed to vary in ellipticity.

\begin{figure}
\hspace{0cm}
\psfig{file=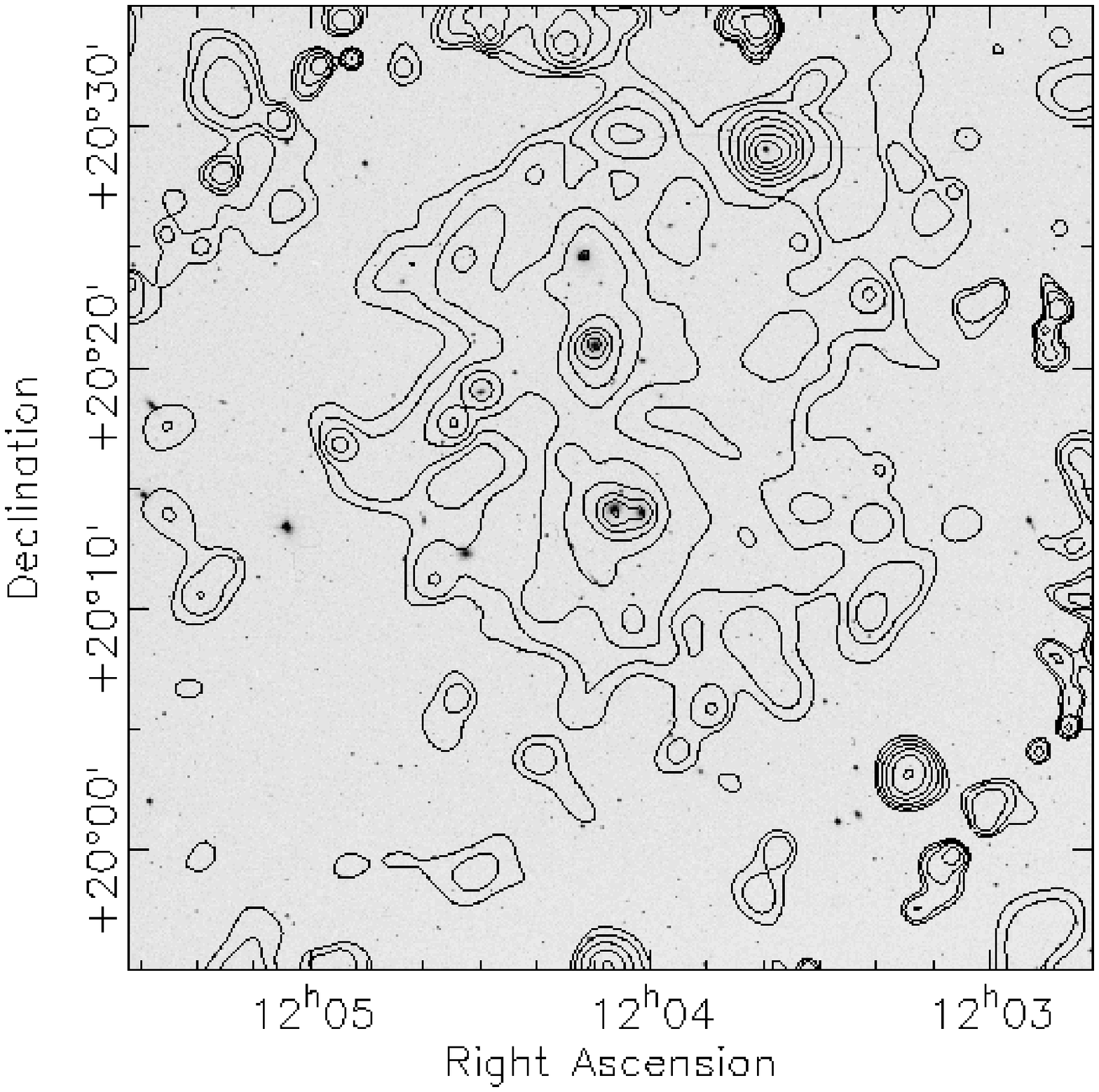,angle=-90,width=8cm}
\caption{\label{fig:ovly1}Contours of adaptively smoothed X-ray emission
  from the group NGC4065, overlaid on an optical image. It is clear that
  there are two distinct centres of emission in this system.}
\end{figure}

In three of the groups (NGC4065, NGC4073 and NGC7619) the emission was
bimodal, so that the two-component models fitted with the centres of the
two components significantly offset from one another (e.g. see
Fig.\ref{fig:ovly1}). As a result, it is not sensible to define one
component as extended, and the other as the central component. In these
cases both of the components were constrained to be spherical.

\begin{table*}
\begin{minipage}[c]{18cm}
  \center{\caption{\label{tab:2kingfit}Results of the surface brightness
      fits for elliptical and two-component models. If two models are shown
      for a group, the first is the elliptical model and the second the the
      two-component model. Models marked with an $\ast$ are groups in which
      two separate centres of emission could be observed. The goodness of
      fit is as described in the main body of the text. All errors are
      $1\sigma$ for one interesting parameter.}}
\begin{tabular}{lcccccccc}
\hline
 & & Extended & component & & & Central & component & \\
\cline{2-5}  \cline{7-8}
Group   & $\beta_{fit}$     & Core radius        & Axis ratio & Position angle & & Core radius & $\beta_{fit}$             & Goodness \\  
 & &(arcmin) & &(degrees) & & & & of fit \\
\hline                                                                          
NGC 315  & 1.37 $\pm$ 0.36   & 0.42 $\pm$ 0.10    & 1.08 $\pm$ 0.13 & 141 $\pm$ 65 & & - & - & -0.5 \\     
NGC 383  & 0.362 $\pm$ 0.003 & 0.43 $\pm$ 0.06    & 1.34 $\pm$ 0.05 & 166 $\pm$ 4  & & - & - & 8.2   \\     
        & 0.48 $\pm$ 0.02   & 6.9 $\pm$ 0.9     & 1.19 $\pm$ 0.04 & 156 $\pm$ 6  & & 0.01 $\pm$ 0.03 & 0.48 $\pm$ 0.02   & 5.2   \\     
NGC 524  & 0.45 $\pm$ 0.01   & 0.01 $\pm$ 0.01  & 1.41 $\pm$ 0.22 & 143 $\pm$ 16 & & - & - & 18.2   \\     
NGC 533  & 0.482 $\pm$ 0.005 & 0.40 $\pm$ 0.03    & 1.60 $\pm$ 0.06 & 77 $\pm$ 3   & & - & - & 14.2   \\                                            
        & 0.75 $\pm$ 0.14   & 10.2 $\pm$ 2.8    & 1.89 $\pm$ 0.04 & 49 $\pm$ 2   & & 0.02 $\pm$ 0.02 & 0.54 $\pm$ 0.02   & -0.23 \\     
NGC 741  & 0.465 $\pm$ 0.008 & 0.37 $\pm$ 0.05    & 1.35 $\pm$ 0.09 & 74 $\pm$ 6   & & - & - & 11.2   \\                                                                                                                             
        & 0.391 $\pm$ 0.009 & 0.10 $\pm$ 0.13   & 1.24 $\pm$ 0.11   & 174 $\pm$ 11 & & 0.17 $\pm$ 0.09 & 0.9 $\pm$ 0.3     & 0.20  \\     
NGC 1587 & 0.47 $\pm$ 0.06   & 0.34 $\pm$ 0.27    & 1.4 $\pm$ 0.5 & 35 $\pm$ 22  & & - & - & -0.1 \\     
NGC 2563 & 0.369 $\pm$ 0.003 & 0.01 $\pm$ 0.02   & 1.28 $\pm$ 0.04 & 88 $\pm$ 7   & & - & - & 4.6   \\                                                                                                                             
        & 0.400 $\pm$ 0.004 & 2.6 $\pm$ 0.6     & 1.31 $\pm$ 0.06 & 95 $\pm$ 7   & & 0.2 Fixed       & 1.0 Fixed         & 1.54  \\     
NGC 3091 & 0.60 $\pm$ 0.02 & 0.48 $\pm$ 0.06    & 1.64 $\pm$ 0.09 & 14 $\pm$ 4   & & - & - & 10.0   \\                                                                                                                             
        & 0.41 $\pm$ 0.02 & 0.1 $\pm$ 0.05   & 4.3 $\pm$ 1.3   & 184 $\pm$ 4  & & 0.3 $\pm$ 0.07 & 0.66 $\pm$ 0.05       & 0.82  \\     
NGC 3607 & 0.52 $\pm$ 0.18 & 4.8 $\pm$ 1.2      & 3.4 $\pm$ 0.7 & 20 $\pm$ 3   & & - & - & 3.2   \\                                                                                                                             
$\ast$  & 0.45 $\pm$ 0.04 & 0.28 $\pm$ 0.16   & 1.0 Fixed         &  0.0 Fixed   & & 0.01 $\pm$ 0.04 & 0.38 $\pm$ 0.03  & 0.63  \\     
NGC 3665 & 0.49 $\pm$ 0.03 & 0.13 $\pm$ 0.12    & 1.4 $\pm$ 0.2 & 63 $\pm$ 19  & & - & - & 0.2  \\     
NGC 4065 & 0.47 $\pm$ 0.04 & 4.1 $\pm$ 0.6      & 3.6 $\pm$ 0.4 & 7 $\pm$ 1    & & - & - & 8.9   \\                                                                                                                             
$\ast$  & 0.41 $\pm$ 0.01 & 0.05 $\pm$ 0.06   & 1.0 Fixed         &  0.0 Fixed   & & 4.3 $\pm$ 1.4   & 0.8 $\pm$ 0.2    & 2.49  \\     
NGC 4073 & 0.431 $\pm$ 0.002 & 0.10 $\pm$ 0.01    & 1.25 $\pm$ 0.02 & 103 $\pm$ 3  & & - & - & 2.5   \\                                                                                                                          
        & 0.46 $\pm$ 0.01 & 2.14 $\pm$ 0.26   & 1.28 $\pm$ 0.04 & 102 $\pm$ 4  & & 0.34 $\pm$ 0.05 & 0.73 $\pm$ 0.06   & 1.64  \\  
NGC 4261 & 0.446 $\pm$ 0.004 & 0.01 $\pm$ 0.01    & 1.34 $\pm$ 0.09 & 47 $\pm$ 5   & & - & - & 13.6   \\                                                                                                                             
        & 0.35 $\pm$ 0.03 & 3.3 $\pm$ 1.0     & 1.22 $\pm$ 0.13   & 156 $\pm$ 17 & & 0.28 $\pm$ 0.01 & 1.0 Fixed         & -1.20 \\     
NGC 4325 & 0.60 $\pm$ 0.01 & 0.32 $\pm$ 0.03    & 1.17 $\pm$ 0.05 & 11 $\pm$ 8   & & - & - & 1.3   \\     
NGC 4636 & 0.476 $\pm$ 0.003 & 0.268 $\pm$ 0.014  & 1.09 $\pm$ 0.08 & 17 $\pm$ 7   & & - & - & 17.0   \\                                                                                                                             
        & 0.373 $\pm$ 0.008 & 0.013 $\pm$ 0.006 & 1.15 $\pm$ 0.06 & 167 $\pm$ 9  & & 0.92 $\pm$ 0.11 & 0.81 $\pm$ 0.06   & 6.39  \\     
NGC 4761 & 0.502 $\pm$ 0.005 & 0.30 $\pm$ 0.02    & 1.33 $\pm$ 0.04 & 16 $\pm$ 3   & & - & - & 19.3   \\                                                                                                                             
        & 0.364 $\pm$ 0.006 & 0.10 $\pm$ 0.04   & 1.17 $\pm$ 0.09 & 72 $\pm$ 14  & & 0.49 $\pm$ 0.06 & 0.85 $\pm$ 0.04   & 1.28  \\     
NGC 5129 & 0.44 $\pm$ 0.01 & 0.15 $\pm$ 0.06    & 1.30 $\pm$ 0.16 & 28 $\pm$ 13  & & - & - & -0.7 \\ 
NGC 5171 & 0.34 $\pm$ 0.03 & 0.84 $\pm$ 0.71    & 2.7 $\pm$ 0.9 & 171 $\pm$ 6  & & - & - & 0.8  \\     
NGC 5353 & 0.58 $\pm$ 0.03 & 1.35 $\pm$ 0.17    & 1.34 $\pm$ 0.11 & 129 $\pm$ 7  & & - & - & 71.0  \\                                                                                                                        
        & 0.44 $\pm$ 0.02 & 0.29 $\pm$ 0.13   & 1.8 $\pm$ 0.2   & 25 $\pm$ 5   & & 0.01 $\pm$ 0.01 & 1.0 Fixed         & 6.39  \\
NGC 5846 & 0.66 $\pm$ 0.02 & 1.43 $\pm$ 0.08    & 1.70 $\pm$ 0.05 & 246 $\pm$ 2  & & - & - & 106.9  \\                                                                                                                         
        & 0.58 $\pm$ 0.01 & 0.84 $\pm$ 0.07   & 1.16 $\pm$ 0.03 & 47 $\pm$ 7   & & 0.27 $\pm$ 0.04 & 1.0 Fixed         & 3.48  \\ 
NGC 6338 & 0.423 $\pm$ 0.004 & 0.06 $\pm$ 0.01    & 1.24 $\pm$ 0.06 & 24 $\pm$ 7   & & - & - & 21.9  \\                                                                                                                             
        & 0.52 $\pm$ 0.04 & 2.6 $\pm$ 0.4   & 1.29 $\pm$ 0.07 & 23 $\pm$ 6   & & 0.42 $\pm$ 0.03 & 1.0 Fixed         & 1.40  \\     
NGC 7176 & 1.07 $\pm$ 0.29   & 8.0 $\pm$ 4.2      & 1.6 $\pm$ 0.2 & 139 $\pm$ 10 & & - & - & 2.3  \\     
NGC 7619 & 0.458 $\pm$ 0.006 & 1.42 $\pm$ 0.08    & 1.93 $\pm$ 0.07 & 210 $\pm$ 1  & & - & - & 22.7  \\                                                                                                                        
$\ast$  & 0.78 $\pm$ 0.08 & 0.26 $\pm$ 0.05   & 1.0 Fixed         & 0.0 Fixed    & & 0.01 $\pm$ 0.01 & 0.40 $\pm$ 0.01  & 6.27  \\
NGC 7777 & 0.35 $\pm$ 0.02 & 0.01 $\pm$ 0.01  & 1.0 Fixed       & 0.0 Fixed    & & - & - & 1.1  \\
\hline
\end{tabular}
\end{minipage}
\end{table*}

The fitted parameters of the two-component King profiles are also shown in
Table~\ref{tab:2kingfit}, along with an estimate of the goodness of fit.
The errors quoted are 1$\sigma$ for one interesting parameter. Note that
these errors are only reliable for reasonable fits (see final column in
Table~\ref{tab:2kingfit}). The best fitting surface brightness profiles
were also used to correct the derived group fluxes for the diffuse emission
lost when point sources are removed. A model image for the group was
produced, and from the ratio of the number of counts in the model image to
that in the same image with `holes' punched at the positions of detected
sources, a correction factor for the fluxes was obtained. The luminosity of
each group was then calculated using distances corrected for infall to
Virgo and the Great Attractor \cite{fixsen96,burstein90}, which are listed
in Table~\ref{tab:groupspec}.

In the case of groups with a detected central component, we checked for the
possibility that this might arise from a nuclear source in the central
galaxy. Fits with Gaussian models for the central component show that it is
extended at $>99$\% confidence in all cases except NGC 5353, where
statistics are too poor to constrain the extent of the central source. For
each of these systems a search for radio sources associated with the
brightest group galaxies was carried out using NED and the \pcite{burns87}
radio survey of groups, which has some overlap with this sample. This
search identified radio sources associated with six of the brightest group
galaxies: NGC 383, NGC 741, NGC 4261, NGC 4636, NGC 5353 and NGC 6338.
\pcite{hwang99} have studied three of these using ASCA spectra.  Two showed
no significant improvement in fit statistic when a powerlaw component was
added to the spectral model.  For the third, NGC 6338, \pcite{hwang99} find
evidence that there may be contamination by an AGN, although in our data
the spatial extent of the central component in this, and almost all the
other systems rules out a large AGN component. The only system that may be
contaminated (as indicated by the spatial extent of the central component)
is NGC 5353. For this system we fitted the spectral data for this group
with an added power-law component of index 1.7. We then calculated the
relative contributions from the power-law and hot plasma components in the
{\it ROSAT} band. This showed that even for the 90\% upper limit of the
power-law component the emission was dominated by the hot plasma component.
Our conclusion from these spatial and spectral studies is that any AGN
contribution to the central components in these systems appears to be
minor.

We were also interested in the way in which the luminosity of groups varies
with radius. The model images were therefore used to calculate luminosities
within radii of 200~kpc, 500~kpc, 1/3 of the virial radius and the virial
radius ($R_V$). Note that $R_V$ lies well beyond the radius to which
significant X-ray emission can actually be detected in our data, in almost
all cases. It can be seen in Table~\ref{tab:2kingfit} that the
two-component models provide good descriptions of the data in the majority
of cases.  However even in the cases where the two-component fit is not
acceptable it is significantly better than the single-component model, thus
where possible, the two-component models are used for the purposes of
calculating the effects of extrapolating to different radii.  The virial
radii of the groups were determined using the relation obtained from
simulations by Navarro, Frenk \& White (1995)\nocite{navarro95a}. This is
given (for H$_0$~=~50~km~s$^{-1}$ Mpc$^{-1}$) by,\\

\noindent
\begin{math}
R_V = 2.57(\frac{T}{5.1KeV})^{\frac{1}{2}} \textrm{ Mpc.}
\end{math}
\\

Luminosities and temperatures derived in this study are generally similar
to those from earlier studies
\cite{doe95,ponman96a,mulchaey96,burns96,mulchaey98b}. The small
differences in the luminosities of groups common to both this sample and
that of \scite{burns96} are most likely primarily due to the fact that
\scite{burns96} use a spectral model with a temperature of 1 keV to derive
all luminosities, whereas the luminosities derived here use fitted spectral
models, and thus should be more reliable.

\begin{table*}
\begin{minipage}[c]{18cm}
  \center{\caption{\label{tab:groupspec}Results of the spectral fitting are
      shown along with the derived luminosities for each of the groups. The
      distance to each of the groups is calculated after allowances for
      infall to Virgo and the Great Attractor. The luminosities are those
      derived within the radius to which emission could be observed. The
      final column shows what fraction of the luminosity as extrapolated to
      the virial luminosity is observed within the stated radius. All
      errors are $1\sigma$.}}
\begin{tabular}{lcccccccc}
\hline
 & & & & & & Radius of & Physical & Fraction of \\
Group name & N$_H$ & Temperature   & Abundance       & Distance & log L           & extraction & radius & virial \\  
 & (1.0$^{21}$ cm$^{-2}$) & (keV) & (Solar) & (Mpc) & (erg s$^{-1}$) & (arcmin) & (kpc) & luminosity \\
\hline                                                                          
NGC 315  & 0.588 & 0.85 $\pm$ 0.07  & 0.12 $\pm$ 0.05 & 96.4 & 42.15 $\pm$ 0.15 & 6.0  & 168  & 1.00$\ast$\\     
NGC 383  & 0.54 & 1.53 $\pm$ 0.07  & 0.40 $\pm$ 0.09 & 101.8 & 43.31 $\pm$ 0.02 & 30.0  & 889  & 0.76\\     
NGC 524  & 0.467 & 0.56 $\pm$ 0.08  & 0.35 $\pm$ 0.43 & 49.9 & 41.37 $\pm$ 0.11 & 10.6  & 153  & 0.59\\     
NGC 533  & 0.305 & 1.06 $\pm$ 0.04  & 0.82 $\pm$ 0.19 & 106.8 & 42.95 $\pm$ 0.02 & 20.3  & 630  & 0.91\\     
NGC 741  & 0.442 & 1.08 $\pm$ 0.06  & 0.48 $\pm$ 0.18 & 106.0 & 42.66 $\pm$ 0.03 & 16.0  & 494  & 0.59\\     
NGC 1587 & 0.692 & 0.92 $\pm$ 0.15  & 0.3 Fixed       & 77.1 & 41.50 $\pm$ 0.18 & 6.0  & 134  & 0.46\\     
NGC 2563 & 0.424 & 1.06 $\pm$ 0.04  & 0.56 $\pm$ 0.14 & 106.7 & 42.79 $\pm$ 0.02 & 17.6  & 546  & 0.57\\     
NGC 3091 & 0.478 & 0.71 $\pm$ 0.03  & 0.95 $\pm$ 1.08 & 90.6 & 42.20 $\pm$ 0.03 & 8.9  & 236  & 0.81\\     
NGC 3607 & 0.156 & 0.41 $\pm$ 0.04  & 0.05 $\pm$ 0.02 & 33.1 & 41.59 $\pm$ 0.03 & 9.6  & 92  & 0.34\\     
NGC 3665 & 0.204 & 0.45 $\pm$ 0.11  & 0.17 $\pm$ 0.14 & 52.5 & 41.36 $\pm$ 0.10 & 6.0  & 92  & 0.66\\     
NGC 4065 & 0.239 & 1.22 $\pm$ 0.08  & 0.80 $\pm$ 0.36 & 151.3 & 42.99 $\pm$ 0.04 & 15.0  & 660 & 0.76\\     
NGC 4073 & 0.190 & 1.59 $\pm$ 0.06  & 0.94 $\pm$ 0.12 & 135.9 & 43.70 $\pm$ 0.01 & 18.0  & 711  & 0.75\\  
NGC 4261 & 0.156 & 0.94 $\pm$ 0.03  & 0.20 $\pm$ 0.02 & 53.0 & 42.32 $\pm$ 0.02 & 15.0  & 231  & 0.26\\     
NGC 4325 & 0.223 & 0.86 $\pm$ 0.03  & 2.01 $\pm$ 1.09 & 162.6 & 43.35 $\pm$ 0.03 & 10.2  & 482  & 0.95\\     
NGC 4636 & 0.179 & 0.72 $\pm$ 0.01  & 0.51 $\pm$ 0.04 & 32.8 & 42.48 $\pm$ 0.01 & 21.6  & 206  & 0.46\\     
NGC 4761 & 0.297 & 1.04 $\pm$ 0.02  & 0.36 $\pm$ 0.04 & 104.7 & 43.16 $\pm$ 0.01 & 15.6  & 475  & 0.59\\     
NGC 5129 & 0.176 & 0.81 $\pm$ 0.06  & 0.51 $\pm$ 0.22 & 149.9 & 42.78 $\pm$ 0.04 & 9.0  & 393 & 0.65\\ 
NGC 5171 & 0.193 & 1.05 $\pm$ 0.11  & 0.40 $\pm$ 0.20 & 150.1 & 42.92 $\pm$ 0.05 & 10.8  & 472 & 0.39\\     
NGC 5353 & 0.0973 & 0.68 $\pm$ 0.05  & 0.30 $\pm$ 0.07 & 58.0 & 41.76 $\pm$ 0.03 & 9.6  & 162 & 0.46\\
NGC 5846 & 0.428 & 0.70 $\pm$ 0.02  & 0.43 $\pm$ 0.10 & 42.3 & 42.36 $\pm$ 0.02 & 15.0  & 185 & 0.88\\ 
NGC 6338 & 0.256 & 1.69 $\pm$ 0.16  & 0.06 $\pm$ 0.04 & 171.2 & 43.93 $\pm$ 0.01 & 13.8  & 687 & 0.79\\     
NGC 7176 & 0.163 & 0.53 $\pm$ 0.11  & 0.3 $\pm$ 0.6 & 50.6 & 41.47 $\pm$ 0.11 & 13.5  & 199 & 0.90\\     
NGC 7619 & 0.496 & 1.00 $\pm$ 0.03  & 0.44 $\pm$ 0.09 & 64.9 & 42.62 $\pm$ 0.02 & 24.0 & 453 & 0.59\\
NGC 7777 & 0.500 & 0.62 $\pm$ 0.15  & 0.3 Fixed       & 133.4 & 41.75 $\pm$ 0.20 & 6.6  & 256 & 0.32\\
\hline                                      
\end{tabular}
\\
$\ast$ NGC 315 fits with a high $\beta_{fit}$ and it is possible that for
this group the emission may be due to a extensive elliptical galaxy halo
rather than genuine group emission.
\end{minipage}
\end{table*}

\section{Results}
\label{sec:Distributionsandcorrelations}

Throughout the following sections, the luminosities quoted are extracted
from within the radius given in Table~\ref{tab:groupspec}.  Corrections for
removed point sources have been made using the best model derived for each
group; either two-component or (elliptical) single-component.

\subsection{X-ray profiles}

\begin{figure}
\hspace{0cm}
\psfig{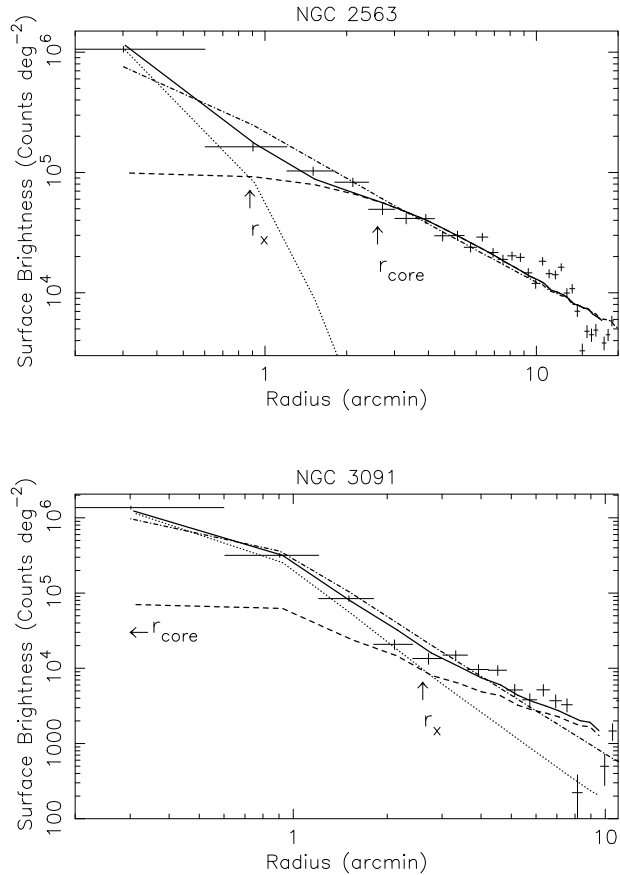}
\caption{\label{fig:sbprofs}1D surface brightness profiles for the groups NGC
  2563 and NGC 3091. The overall best fit two component models are shown as
  the solid lines, with the dotted line representing the central component
  and the dashed line the extended. Data points are shown as crosses. For
  comparison the single component elliptical models are marked as the
  dot-dashed lines. This is steeper than the extended component in NGC 3091
  but slightly flatter than it in NGC 2563. $r_{core}$ marks the core
  radius of the extended component and $r_x$ the crossover radius, as
  defined in the text.}
\end{figure}

The surface brightness profiles for our 24 systems break down into 12
two-component, 9 single-component and 3 bimodal cases. However, note that
the nine single-component systems include the eight groups with the lowest
source counts in the sample, so it is likely that the majority of these
single-component fits appear to be adequate only because of poor
statistics. Two examples of radial profiles are shown in
Fig.\ref{fig:sbprofs}. These 1D profiles only give an approximate
representation of our 2D models, but the centres of the two components
almost coincide in the two cases shown, and profiles for both data and
model components have been derived about the centre of the more compact
component. A distinct `shoulder' in the observed profile indicates the need
for two components in the model, as noted by MZ98.

The median value of $\beta_{fit}$ obtained for the extended group component
in our sample (from two-component fits where available, or else
single-component fits), is $\beta_{fit}=0.46$, and the weighted mean is
$0.42\pm0.06$ (where the error is derived from the scatter of the values
about the mean). This value of $\beta_{fit}$ can be compared with the
typical value for rich clusters, $\beta_{fit}\approx 2/3$
(\pcite{arnaud99}; Mohr, Mathiesen \& Evrard 1999)\nocite{mohr99}),
indicating that the surface brightness profiles of the groups in this
sample are generally significantly flatter than those of clusters.

In the case of a number of the groups, the core radius derived for the
extended group component is smaller than the resolution of the {\it ROSAT}
PSPC.  Such a small core radius means essentially that the group emission
has been fitted with a power law model. In such cases, the derived core
radii are unreliable, particularly if an additional central component is
present. To investigate the effect of a small core radius on the other
derived parameters, in particular the slope, $\beta_{fit}$, we varied the
core radii in a number of groups which fitted with two components and a
small core radius for the extended component. It was found that varying the
core radii between 0.1 and 1.0 arcmin typically changes the value of
$\beta_{fit}$ by less than 5\%.  Values of the index and core radius of the
central component also varied only within 1$\sigma$ of their best fit
values. Hence uncertainties in core radius in such cases do not seriously
compromise our results for other parameters.

We define the `cross-over radius', $r_x$, to be the radius within which the
central component dominates the surface brightness. We derived values for
$r_x$ for the twelve groups for which two-component fits were available,
but the emission was not bimodal. The mean cross-over radius for these
systems was $r_x=35 \pm 6$~kpc. Groups with two-component profiles which
have $r_{core}<r_x$ are deemed to have poorly determined core radii. In
order to gain some insight into the typical core radii of galaxy groups,
the median value was determined, using $r_x$ as an upper limit for those
groups with $r_{core}<r_x$. Under these assumptions the median core radius
of the twelve groups was found to be 60 kpc.

\begin{figure*}
\hspace{0cm}
\psfig{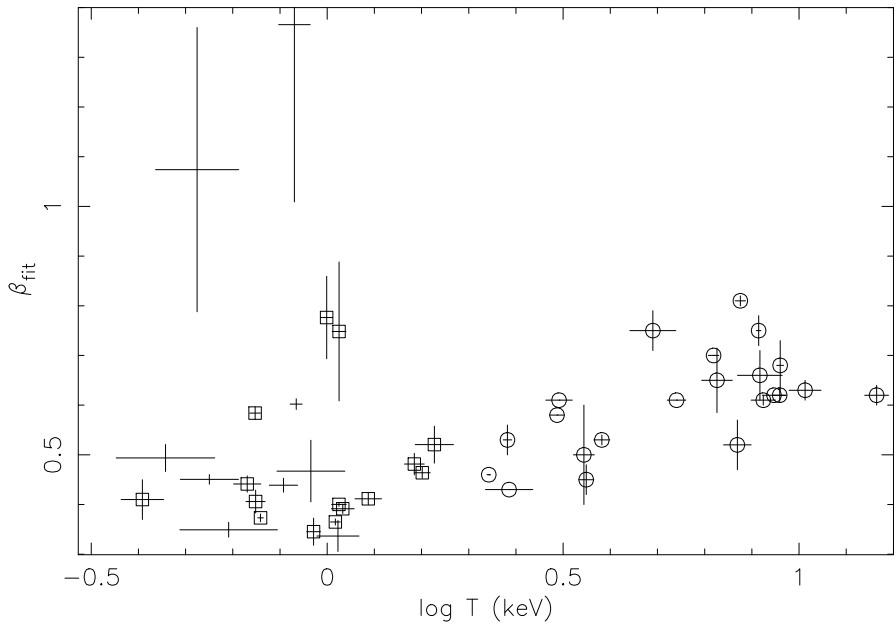}
\caption{\label{fig:beta_temp_1}The relationship between
  $\beta_{fit}$ and temperature for the whole group sample compared to
  cluster data from Arnaud \& Evrard (1999). Values from groups with
  single-component fits are shown as plain crosses, those from
  two-component fits as crosses with central squares. The cluster data are
  marked as crosses with central circles.}
\end{figure*}
\nocite{arnaud99}

\begin{figure*}
\hspace{0cm}
\psfig{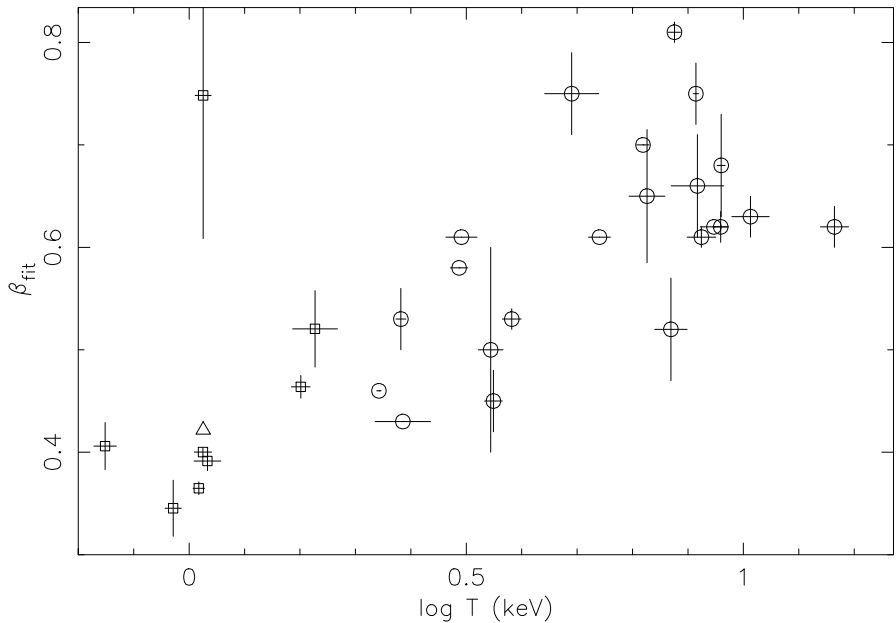}
\caption{\label{fig:beta_temp_2}The relation between $\beta_{fit}$
  and temperature for a subsample of the groups with the best two-component
  models, and the cluster data. Symbols are the same as in the previous
  figure. The triangle marks the new value of the one discrepant group
  value when its central component was refitted with a Gaussian.}
\end{figure*}

The relationship between the integrated temperature of the intragroup gas
and the best obtained $\beta_{fit}$ value for each of the groups is plotted
in Fig.\ref{fig:beta_temp_1}. Also shown are cluster data from
\scite{arnaud99} (data points with circles). The group data are split into
two categories: single-component (plain crosses) and the extended component
from the two-component fits (points with square in centre). As can be seen,
the general trend in clusters is for $\beta_{fit}$ to drop with decreasing
temperature. In the region of the graph containing the group data it is
clear that the majority of the groups have low $\beta_{fit}$ values, but
there is also a large amount of scatter, in particular amongst the groups
with single-component fits.

Fig.\ref{fig:beta_temp_1} also includes the three bimodal groups.
Excluding these three groups, the single-component fits and the
two-component fits whose quality of fit (from the Monte Carlo simulations)
is poor, greatly reduces the scatter in the group results. The outcome is a
much clearer trend in $\beta_{fit}$ with temperature, as can be seen in
Fig.\ref{fig:beta_temp_2}. The combined groups and cluster data are
significantly correlated with a Kendall's rank correlation coefficient (a
distribution free test for correlation) of K=4.05 (P=0.00006 of chance
occurrence). The one group point that conflicts with the general trend
(NGC533) is, in fact, the only group in the sample which has a flatter
$\beta_{fit}$ value for its central component than for the extended
component. This means that the central component has a significant effect
beyond the central region. Hence the {\it shape} of this component could
affect the parameters obtained for the extended component. To test this, we
refitted the surface brightness profile with a Gaussian model for the
central component, in place of the previous King model. The $\beta_{fit}$
value for the extended component changed markedly, and the new value is
denoted by the triangle in Fig.\ref{fig:beta_temp_2}. As can be seen, this
point is now much closer to the trend described by the other groups.

\subsection{Luminosity, temperature and velocity dispersion}

\subsubsection{X-ray luminosities}

Bolometric luminosities for each group, derived from within the extraction
radius as described in section~3, are given in Table~\ref{tab:groupspec},
along with best fit spectral properties. The tabulated luminosities are
those of the intragroup gas only. Errors on the luminosities are derived
from Poisson errors on the data combined with errors arising from
uncertainties in the fitted surface brightness profiles, which are used to
correct for flux lost where contaminating sources have been excluded.

The flat surface brightness profiles of groups imply that a significant
fraction of their luminosity derives from large radii. To quantify this, we
used our best fit surface brightness models to derive bolometric
luminosities extrapolated to $R_V$, and the fraction of this luminosity
represented by the luminosity derived from within the extraction radius is
shown for each group in Table~\ref{tab:groupspec}. This may be as low as
$\sim$30\% in some cases.

\begin{figure}
\hspace{0cm}
\psfig{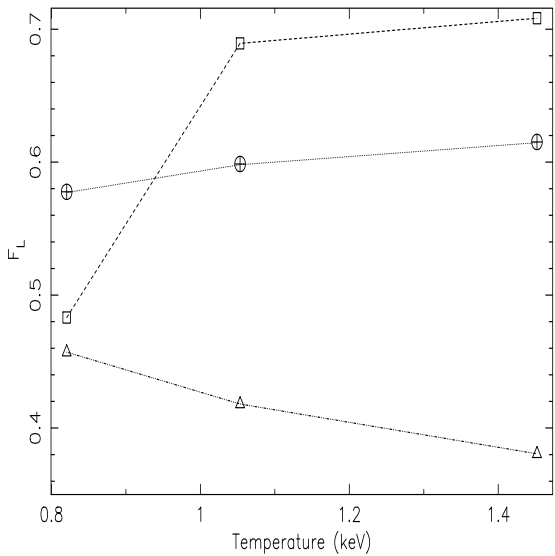}
\caption{\label{fig:fraction}Fraction of the total
  luminosity observed within three different radii, as deduced from best
  fitting surface brightness models, for systems of different temperature.
  Squares show the luminosity within the radius to which emission could be
  observed, circles the luminosity within $R_V/3$, and triangles the
  luminosity within a fixed radius of 200~kpc.}
\end{figure}

The effect of scaling the luminosities to different radii is shown in more
detail in Fig.\ref{fig:fraction}. This analysis is based on the eight
systems with well-fitting two-component profiles. These have been binned
into three temperature bins to reduce fluctuations from system to system
and show trends more clearly. The luminosity as a fraction of that within
$R_V$ is shown at three radii for the systems within each temperature bin.
Points marked by triangles (dash-dot-dot line) show these ratios at a
radius of 200~kpc, squares (dashed line) at the radius out to which
emission could be detected, and crossed circles (solid line) at one third
of the virial radius of the group.  As can be seen the luminosity is
significantly underestimated in all cases. In particular, for groups
measured to a fixed radius of 200~kpc, and for the coolest groups at the
extraction radius, one may underestimate $L_X$ by factor of more than two.

\subsubsection{Correlations}

\begin{figure*}
\hspace{0cm}
\psfig{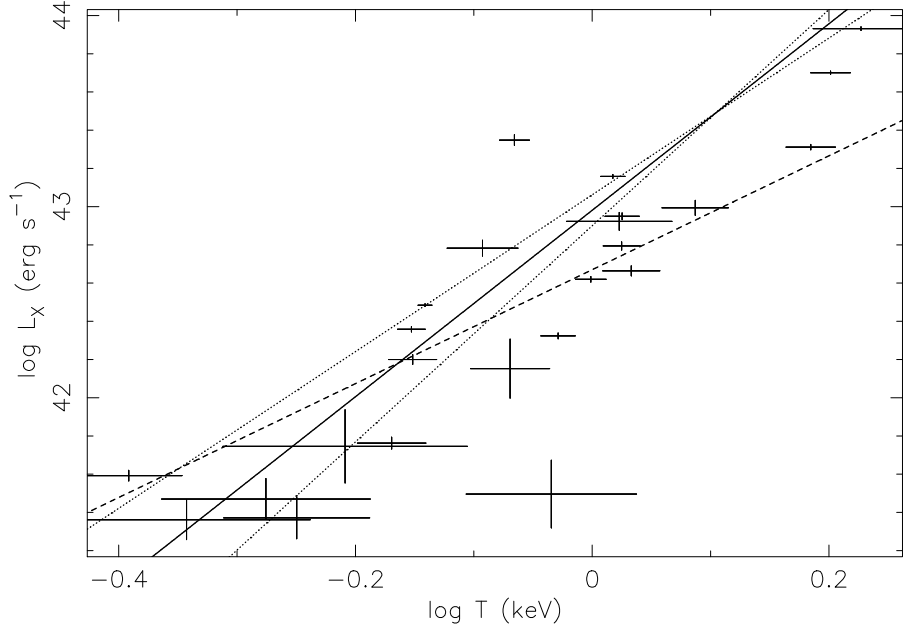}
\caption{\label{fig:L:T}The relation between X-ray
  luminosity and temperature for our group sample. The solid line shows the
  best fit relation to our data, with one sigma error bounds marked by
  dotted lines. The extrapolation of the best fitting cluster relation
  (White et al. 1997) is shown as the dashed line.}
\end{figure*}

The well-known relation between X-ray luminosity and temperature is
apparent in our sample. The two parameters are significantly correlated
(K=4.81, P$<$0.00001) and the relation between them is shown in
Fig.\ref{fig:L:T}. Neither the errors on $L_X$ or $T$ are negligible, and a
doubly weighted technique made available through the \textsc{odrpack}
package was used in this and following plots to determine the best fit
line,\\

\noindent
\begin{math}
\log L_X = (42.98 \pm 0.08) + (4.9 \pm 0.8) \log T
\end{math}
.
\\

This relationship is marked with its 1$\sigma$ error bounds in
Fig.\ref{fig:L:T}. A best fit to the cluster $L:T$ relation has been
derived by White, Jones \& Forman (1997)\nocite{white97}. They obtain $\log
L_X = 42.67 + 2.98 \log T$, which is marked as the heavy dashed line in
Fig.\ref{fig:L:T}. This line is much flatter than the best trend fit for
the loose groups.

The luminosities used in this plot are those within the radius of
extraction. Fig.\ref{fig:fraction} shows that at this radius the
luminosities will be underestimated, with the effect being greatest in the
smaller mass systems. This means that if luminosities extrapolated to the
virial radius were used, the $L:T$ slope should be slightly flatter. This
is indeed found to be the case, with a best fitting relation of $\log L_X =
(43.17 \pm 0.07) + (4.2 \pm 0.7) \log T$, although the difference in slope
from the previous relation is not formally significant.

\nocite{white97}

If galaxy systems scaled with mass in a self-similar way, then one would
expect $L_X \propto T^2$.  The cluster relation is steeper than this, and
our result for groups is steeper still. However, the relationships derived
by \scite{white97} do not take into account the effects of cluster cooling
flows, and recent work suggests that the $L:T$ relation may be flattened
towards $L_X \propto T^2$ when the effects of cooling flows are allowed for
\cite{allen98,markevitch98}.  Such a flattening of the relation for
clusters would raise its extrapolation at low temperatures, accentuating
the disagreement with the low luminosities observed in groups.

\begin{figure*}
\hspace{0cm}
\psfig{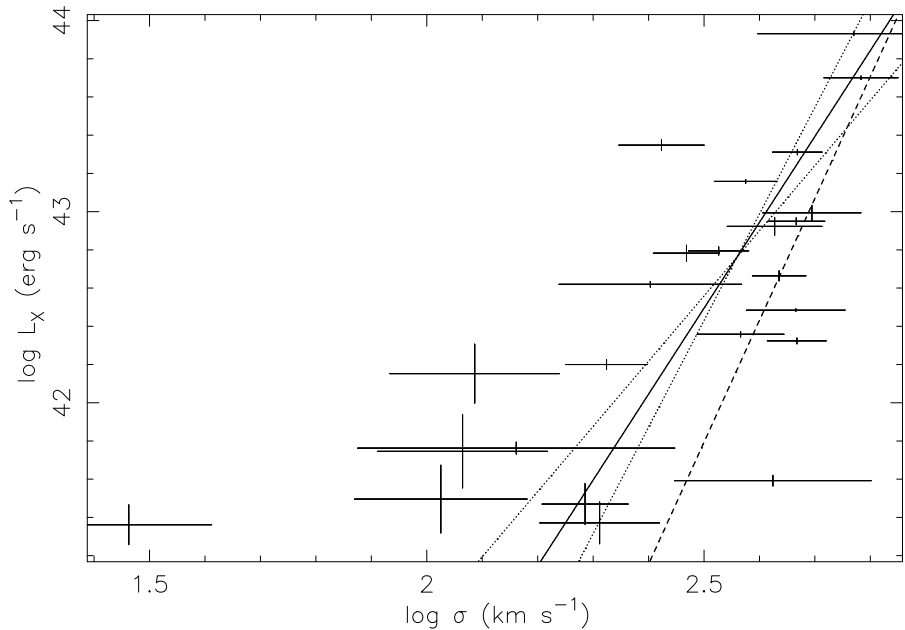}
\caption{\label{fig:L:v}The relationship between X-ray luminosity and group
  velocity dispersion, $\sigma$. The best fit to the data is shown as the
  solid line with the one sigma error bounds marked by the dotted lines.
  The extrapolation of the cluster relation (White et al. 1997) is shown as the
  dashed line.}
\end{figure*}
\nocite{white97}

In Fig.\ref{fig:L:v}, velocity dispersion is plotted against the X-ray
luminosity for our sample. A strong correlation can be seen between these
two parameters (K=3.97, P=0.00006). A regression line fitted to the data
gives \\

\noindent
\begin{math}
\log L_X = (31.3 \pm 2.8) + (4.5 \pm 1.1) \log \sigma
\end{math}
,
\\

\noindent which is marked in Fig.\ref{fig:L:v} with its 1$\sigma$ error
bounds. This relationship is somewhat flatter than the cluster trend given
by \scite{white97} of $\log L_X = 25.84 + 6.38 \log \sigma$ (bold dashed
line in Fig.\ref{fig:L:v}).  Dell'Antonio, Geller \& Fabricant
(1994)\nocite{dellantonio94} found evidence that the $L:\sigma$ relation
may flatten below $\sigma \approx$ 300 km s$^{-1}$.  However they did not
remove the galaxy contribution from the X-ray emission, and suggest that
their flattening may arise from the galaxy contribution becoming
significant at low luminosities.  This flattening has also been confirmed
by \scite{mahdavi97}. In the work presented here, contaminating sources
were removed, but a flatter relation than clusters is still seen. Our
result is actually consistent with that expected from self similar scaling
of clusters, i.e. $L_X \propto \sigma^4$.  However, errors are large and
there is a good deal of scatter, so that the disagreement with the cluster
result is not highly significant, and requires further confirmation.

\begin{figure*}
\hspace{0cm}
\psfig{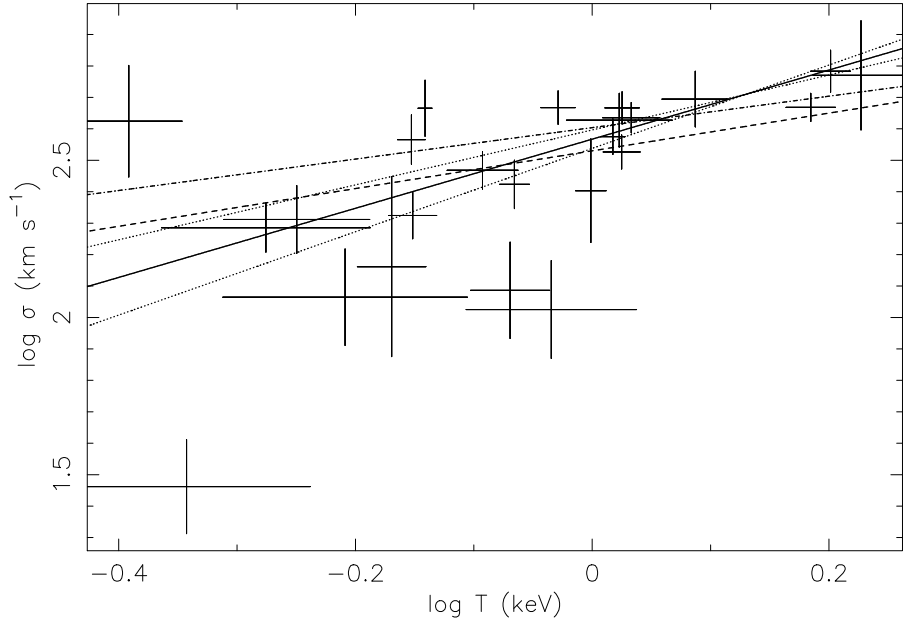}
\caption{\label{fig:v:T}The relationship between group velocity dispersion,
  $\sigma$ and temperature. The best fit to the data is shown as the solid
  line with the one sigma error bounds marked by the dotted lines. The
  extrapolation of the cluster relation (White et al. 1997) is shown as the
  dashed line. The dot-dashed line shows the locus along which
  $\beta_{spec}=1$.}
\end{figure*}
\nocite{white97}

A strong correlation between $\sigma$ and $T$ is shown in Fig.\ref{fig:v:T}
(K=3.82, P=0.0001). A regression line fitted to the data gives \\

\noindent
\begin{math}
\log \sigma = (2.57 \pm 0.03) + (1.1 \pm 0.2) \log T
\end{math}
,\\

\noindent which is shown in Fig.\ref{fig:v:T} with its 1$\sigma$ error
bounds. Also shown in Fig.\ref{fig:v:T} is the line $\beta_{spec}=1$, where
$\beta_{spec}$ is defined as the ratio of the specific energy in the
galaxies to that in the gas. As can be seen, this $\beta_{spec}=1$ line is
flatter than the relation for the loose group sample. However it is
interesting to note that the higher temperature groups appear to be
consistent with $\beta_{spec}=1$, while the lower temperature groups appear
to fall well below this relation. The extension of the best fit relation
for galaxy clusters as determined by \scite{white97} is shown as the dashed
line in Fig.\ref{fig:v:T}. This line, given by $\log \sigma = 2.53 + 0.6
\log T$, is also significantly flatter than the relation determined for the
loose group sample.

The unweighted mean value of $\beta_{spec}$ for our sample is $0.86 \pm
0.13$. However, with one exception, it is clear that $\beta_{spec}$ is
decreasing in the lower temperature (i.e. lower mass) systems.  These
results are in good agreement with those of Bird, Mushotzky \& Metzler
(1995)\nocite{bird95}, who predict a trend towards lower $\beta_{spec}$ in
smaller systems.

The one low temperature point (NGC 3607) that has a high velocity
dispersion is also deviant in the $L:\sigma$ plot. Examination of the group
members reveals that, of the three catalogued members, one is a large
angular distance from the remaining two, and has a large difference in
recession velocity. Also there is a further bright galaxy at the redshift
of the group, which is very close to two of the catalogued members. The
recession velocity of this galaxy is between those of the two catalogued
galaxies, and is almost certainly a group member, although it was not
classified as such by \scite{nolthenius93}. These two effects combined
indicate that the true velocity dispersion of the group is probably
considerably lower than our estimate, which is taken from
\scite{nolthenius93}.

\section{Comparison with Mulchaey and Zabludoff}
\label{Comparison}
As discussed in the introduction, we have included the X-ray bright systems
studied by MZ98, in order to allow a direct comparison of our results with
theirs.  This is important, since our conclusions about $\beta_{fit}$,
$\beta_{spec}$ and the $L:T$ relation all differ from MZ98.  In
Table~\ref{tab:MZgroups} we show the best fit parameters as determined by
MZ98 for the groups that both they and we fit with two-component models
(Note that they also fit two component models to NGC~4325 and NGC~5129,
whilst we find that single component elliptical models provide an adequate
representation of our data for these systems). Whilst we confirm their
conclusion that two-component fits are required to adequately represent
most systems, it can be seen there are some significant differences between
the two sets of results.

\begin{table*}
\begin{minipage}[c]{13cm}
  \center{\caption{\label{tab:MZgroups}A comparison of the two-component
      models fitted by Mulchaey \& Zabludoff (1998) with those from this
      work.  $\beta_{fit}$ and core radius values of the extended component
      for both sets of models are listed. The final column gives the
      difference in Cash statistic between the two models as fitted to our
      data; the negative sign indicating that the model fitted here gives
      the better fit.}}
\begin{tabular}{lccccl}
\hline
 & M\&Z & M\&Z & This work & This work & \\
Group   & $\beta_{fit}$ & core radius  & $\beta_{fit}$ & core radius & $\Delta$C\\  
 &  & (arcmin) &  & (arcmin) & \\
\hline                                                                          
NGC 533  & 0.83     &   8.15       &  0.74 & 10.2  & -122.5 \\     
NGC 741  & 1.00     &   14.08      &  0.39 & 0.1  & -45.1 \\     
NGC 2563 & 0.86     &   11.15      &  0.40 & 2.6  & -26.4 \\     
NGC 3091 & 0.68     &   3.61       &  0.41 & 0.1  & -41.0 \\     
NGC 4761 & 0.63     &   9.00       &  0.36 & 0.1  & -129.2\\     
NGC 5846 & 0.83     &   13.93      &  0.58 & 0.84  & -263.7\\ 
\hline
\end{tabular}
\end{minipage}
\end{table*}
\nocite{mulchaey98b}

The fitting techniques used by MZ98 differ from those used in this work.
Since they work with radial profiles, their fits are necessarily 1D models,
with both components centred at the same point. Their method firstly
involved excluding the central region and fitting for the outer component
only. The central component was then fitted with the extended component
fixed at the values derived from the previous fit. Thus at no stage were
the two components allowed to fit simultaneously. The 2D models fitted in
this work allow the positions of the two components to vary and also permit
elliptical models to be used. Parameters for the two components were also
optimised simultaneously. The lower number of counts in each bin forced us
to use maximum likelihood fitting rather than $\chi^2$ fitting, but the
quality of the fits were checked using the Monte Carlo approach as
described above.

To demonstrate the dangers of a 1D approach to fitting the surface
brightness profiles we simulated an image of a group, in which the outer
component was elliptical (axis ratio=1.5), and offset a short distance (3
arcmin) from the central component. These values were chosen to construct a
fairly elongated and offset system to make any biases more obvious. A 2D
fit successfully recovered the slope of the outer component
($\beta_{fit}$=0.4). We then attempted to fit the data using a 1D approach.
We initially extracted a profile centered on the brightest point in the
group (the central component). This gave a profile with a shoulder and a
clear central excess. This profile was fitted using \textsc{qdp} with a
$\beta$-profile plus a constant background. Initially we fitted to the full
profile, giving a value of $\beta_{fit}\approx0.7$. We then progressively
excluded the central regions and refitted the data. The fitted value of
$\beta_{fit}$ rose to a peak of $\approx0.9$ before dropping as a larger
central region was excluded. Thus it is possible, with the 1D approach used
by MZ98, to significantly overestimate the true value of $\beta_{fit}$.

To decide whether the models of MZ98 referred to in
Table~\ref{tab:MZgroups} provide an acceptable fit to our data, we carried
out a series of two-component fits with the index and core radius frozen at
the MZ98 values. The components were also constrained to be circular and
centred in the same place. The Cash values for these models were then
compared to the best fitting values derived earlier. The differences
between the Cash statistic values are shown in the final column of
Table~\ref{tab:MZgroups}. As can be seen, the models using the MZ98
parameters generally fall well outside the 99\% confidence regions of our
best fitting models (which corresponds to $\Delta$C=-20.1). Hence it
appears that our more sophisticated models do represent the data
significantly better.

The most important difference in the surface brightness results is apparent
in the $\beta_{fit}$ value of the extended component.  MZ98 obtain values
consistent with $\beta_{fit} \approx 1$ whereas the values obtained here
mostly lie in the region 0.4-0.5, with a median value for the extended
component of $\beta_{fit}=0.46$.

MZ98 obtain lower values of $\beta_{fit}$ when fitting single-component
models, but find that the extended components fit with systematically
higher $\beta_{fit}$ when a second component is included (this sort of
effect was reproduced in our simulations mentioned earlier). The same
effect is noted for a sample of clusters by \scite{mohr99}, who give a
useful discussion of the effect. Since core radius and $\beta_{fit}$ are
strongly positively correlated when fitting (i.e. models with larger cores
and higher $\beta_{fit}$ can give rather similar profiles to those with
lower values of both parameters), the presence of a central excess will
force $r_{core}$ towards lower values and hence decrease $\beta_{fit}$,
unless an additional component is included in the model to account for the
central excess.

Interestingly, we do not find this to be the case in general, for our
analysis. For the subset of our groups with two-component fits, the median
value of $\beta_{fit}$ for the single-component fits is 0.47 (i.e. just
steeper than for the two-component fits).  Individually, some groups (e.g.
NGC533) have a steeper profile when the two-component model is used, and
some (e.g.  NGC4761) have a flatter profile. The distinction appears to be
that the argument of \scite{mohr99} applies to systems for which the {\it
extended} component dominates over most of the range of the fitted data.
In this case, the presence of a central component acts to slightly modify
the extended component fit, by reducing both $r_{core}$ and $\beta_{fit}$.
NGC2563 in Fig.\ref{fig:sbprofs} is such an example.  However, for systems
where the {\it central} component is more dominant, such as NGC3091 in
Fig.\ref{fig:sbprofs}, the single component fit is a compromise between a
steeper central component, and a flatter extended one, so that the result
is to {\it increase} $\beta_{fit}$, relative to the extended component.

\begin{figure}
\hspace{0cm}
\psfig{file=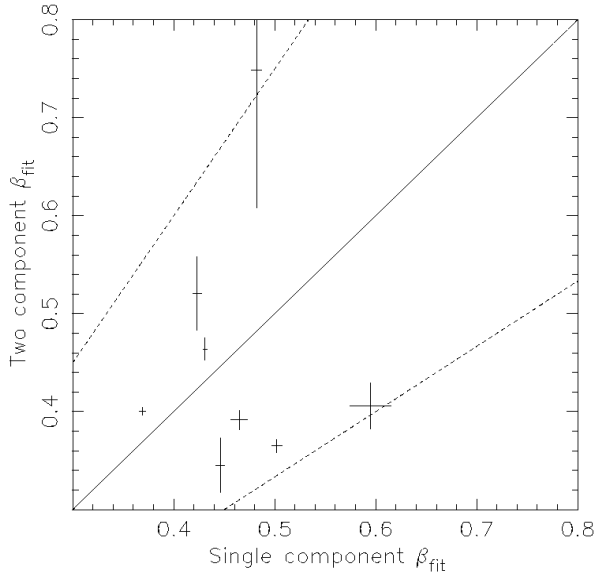,angle=0,height=8cm,width=8cm}
\caption{\label{fig:betacomp}Relationship between the single-component
  $\beta_{fit}$ value and the extended component $\beta_{fit}$ value from
  the two-component fit. The groups shown are the eight systems with the
  best fitting two-component models.}
\end{figure}

Fig.\ref{fig:betacomp} shows the relationship between the $\beta_{fit}$
values from one and two-component models for the eight systems from our
sample with well-fitting two-component profiles. The solid line splits the
graph into two areas. In the upper left area the two-component fit has a
steeper profile than the single-component fit, in the lower right area the
reverse is true.  As can be seen, the single-component fits lead to
overestimates and underestimates of $\beta_{fit}$, relative to the
two-component results, in equal numbers of cases. The two dashed lines
delineate the region in which the two-component fit differs from the
single-component fit by less than $\pm50\%$. As can be seen the
two-component models generally have $\beta_{fit}$ values for the extended
component within 50\% of the single-component fit. The small nominal errors
on the single-component $\beta_{fit}$ values in the figure are misleading,
since they result from calculating errors on a poor fit.

The slope of the $L:T$ relation for our group sample is significantly
steeper than the cluster relation. This is in contrast to the results of
MZ98, who find that the $L:T$ relation for their sample of nine groups is
consistent with the cluster relation. However they had too few points to
fit to the group sample alone, so they added a large cluster sample in
order to determine the best fit line. If the $L:T$ relation is turning over
at a temperature of $\approx$ 1 keV, as is suggested by Fig.\ref{fig:L:T},
then it is to be expected that the line fitted through a combined group and
cluster sample would not differ greatly from the cluster relation.

Values of $\beta_{spec}$ derived by MZ98 for their groups lead them to
conclude that $\beta_{spec} \sim 1$, whereas we see evidence for a drop in
$\beta_{spec}$ for low temperature systems (Fig.~\ref{fig:v:T}). This
difference appears to result from two factors. Firstly, four of the nine
common groups are found in the region ($T$\gtsim 1~keV) where our groups
are generally consistent with $\beta_{spec} \sim 1$. So this only leaves
five systems in which MZ98 could have noted a drop in $\beta_{spec}$.
Secondly, our values of $\beta_{spec}$ appear to be typically about 10\%
lower than those of MZ98. For the nine groups in common, we derive a mean
value of $<$$\beta_{spec}$$>$~=~0.78 compared to
$<$$\beta_{spec}$$>$~=~0.87 for MZ98.  Since we use the same velocity
dispersions, the difference results from the derived gas temperatures. This
difference may arise from the fact that for most groups MZ98 extract their
spectral data from within a larger radius, and given the tendency towards a
decline in temperature with radius apparent in many systems in
Fig.\ref{fig:profiles}, this should result in temperatures somewhat lower
than ours. This interpretation is supported by the fact that our
temperatures are in good agreement with those derived in the study of
\scite{mulchaey96}, in which similar extraction radii were used for systems
common to the two studies.

\section{Discussion}
\label{Discussion}
This survey of X-ray bright, loose galaxy groups represents the largest
detailed study of their properties to date. This allows a comparison with
the properties of richer clusters, and we have been able to show that three
effects are apparent in low temperature systems: steepening of the $L:T$
relation, steepening of the $\sigma$:T relation (i.e. lower $\beta_{spec}$
values in groups), and flatter surface brightness profiles in groups.  We
find that the contrary results of MZ98 appear to be due to the small size
of their sample, coupled with their somewhat less sophisticated analysis of
the surface brightness distributions.

The general nature of these three departures from cluster trends are in
good agreement with the expectations from preheating models, in which
energetic winds from forming galaxies raise the entropy of intergalactic
gas and inhibit its collapse into the smaller potential wells of galaxy
groups (\pcite{metzler94}; Cavaliere, Menci \& Tozzi 1997; Cavaliere, Menci
\& Tozzi 1999; \nocite{cavaliere97,cavaliere99}\pcite{ponman99,metzler99};
Balogh, Babul \& Patton 1999\nocite{balogh99}).  This increase in gas
entropy primarily acts to reduce the gas density in the central regions of
low mass systems, flattening their surface brightness profiles and reducing
their X-ray luminosity. The enhanced entropy also leads to some increase in
gas temperature, resulting in a value of $\beta_{spec}$ less than unity.

The slope of the $L:T$ relation, $L\propto T^{4.9\pm0.8}$, is flatter than
the index of $8.2\pm2.7$ derived for Hickson groups by \scite{ponman96a},
however the error from the HCG sample was very large, so the difference in
slopes is not significant (1.2$\sigma$).  The present, much more accurate
determination of the $L:T$ slope, is in excellent agreement with the
asymptotic relation $L\propto T^5$ derived in the low temperature limit by
the semi-analytical models of \scite{cavaliere97} and \scite{balogh99}.
These two treatments make somewhat different simplifying assumptions about
the physics of the heating of the intracluster gas, but both obtain similar
slopes in the limit of isentropic gas (i.e. where shock heating becomes
negligible).

This result has to be quite robust to detailed model assumptions, since an
approximate result $L \propto T^{4.5} \Lambda(T)$, where $\Lambda(T)$ is
the cooling function, is easily derived by combining the scaling relations
$T\propto M/R$ (from hydrostatic equilibrium), $M\propto R^3$ (for systems
virialising at a given epoch), $\rho_{gas}\propto T^{3/2}$ (for constant
entropy gas) and $L\propto \rho_{gas}^2 \Lambda(T) R^3$. For
bremsstrahlung, $\Lambda(T)\propto T^{1/2}$, so that one obtains $L \propto
T^5$. In practice, at $T\sim 1$~keV the cooling function is flatter than
$T^{1/2}$, due to the increasing contribution of metal lines at low
temperatures, and so the expected relation flattens somewhat towards
$L\propto T^{4.5}$.  The good agreement between this isentropic result and
our observations lends strong support to the result of \scite{ponman99},
that the gas entropy tends towards a constant `floor' value, set by
preheating, in low temperature systems.

Within the above picture, the significant scatter seen in our $L:T$
relation is expected to be primarily due to different star formation and
merging histories of the groups. It has also been shown \cite{fabian94}
that scatter in the cluster $L:T$ relation is correlated with the strength
of the emission associated with a cooling flow. Lower temperature gas (at a
given density) has a shorter cooling time, and it is apparent from
Fig.\ref{fig:profiles} that many of these groups do contain cooling flows.
Hence some $L:T$ scatter can also be attributed to the presence of cooling
flows in the sample.

Another consequence of the effect of galaxy winds is that if winds have
injected extra energy into the intragroup medium then a greater proportion
of the energy of the system should be found in this hot gas. However, this
extra energy could manifest itself in the form of extra thermal energy, or
higher gravitational potential energy of the gas. The models of
\scite{cavaliere99} and \scite{balogh99}, and the N-body+hydrodynamical
simulations of \scite{metzler99}, all indicate that for systems with
$T>1$~keV, the energy is taken up in flattening the gas distribution, with
very little effect on gas temperature. Unfortunately, the simulations of
\scite{metzler99} do not extend to lower temperatures, but the models of
\scite{cavaliere99} and \scite{balogh99} both predict that at
T\ltsim$0.8$~keV, systems depart rather suddenly from the cluster $M:T$
relation, with $T$ flattening out at a minimum value.  This must
necessarily happen, since (in the absence of significant cooling) the gas
temperature cannot drop below the level to which it was preheated, since
its density will have increased as it settles into the group potential.

The observed $\sigma$:T relation for our groups (Fig.\ref{fig:v:T}), is
noisy, but there is a rather clear pattern whereby $\beta_{spec}\approx 1$
for $T$\gtsim 1~keV, but drops to lower values for cooler systems. For
example, the median $\beta_{spec}$ for our nine groups with $T<0.8$~keV is
0.44.  This behaviour is just what the models predict for preheating
temperatures $\sim 0.5$~keV.

The $L:\sigma$ relation for our group sample is slightly flatter than the
cluster relation as determined by \scite{white97}, although the errors on
the slope of the loose group sample are large and as a result the
difference is not statistically significant. This might suggest that the
group L:$\sigma$ relation is an extension of the cluster trend.  However,
if as argued above, preheating has substantially reduced the luminosity of
the groups, then the velocity dispersion must also be lower than expected,
otherwise a steepening of the $L:\sigma$ relation, similar to that seen in
$L:T$, would be observed.

\scite{bird95} have suggested that velocity dispersion should be reduced
for lower mass systems due to the effects of dynamical friction, which is
more effective in lower mass systems due to their lower velocity
dispersion. Loss of orbital energy will lead to a reduction in orbital
velocity provided that the potential is less steep than a singular
isothermal potential in the inner regions. This would be the case for
either a King-like potential, with a flat core, or for potentials of the
form introduced by Navarro, Frenk \& White (1997)\nocite{navarro97}, which
tend to $\rho \propto r^{-1}$ at small radii. However it must be remembered
that the velocity dispersions of the groups in this sample are drawn from
three different sources, and may be based on only a small number of group
galaxies, so that statistical errors are large. \scite{mulchaey98a} find
that when they add the velocities of fainter group galaxies to their
redshift samples, the velocity dispersions they derive may increase by a
factor of 1.5 or more.  This is qualitatively consistent with expectations
from dynamical friction, since the orbits of more massive galaxies should
decay more quickly, and hence their velocity dispersion would drop below
that of fainter group members.

The results on the asymptotic slope of the X-ray surface brightness in
groups derived here, confirms and quantifies the result of
\scite{ponman99}, who showed that surface brightness is progressively
flattened in low temperature systems. This trend is in accord with
preheating models, as discussed above, although our median value of
$\beta_{fit}=0.46$ is a little lower than the values $\beta_{fit}\approx
0.5$-0.6 predicted by the models of \scite{metzler99} and
\scite{cavaliere99} for $T\sim 1$~keV.

The situation in clusters is still a matter of debate.  \scite{arnaud99}
collect together results from the literature, and find a clear trend in
$\beta_{fit}$ with temperature, as can be seen in
Fig.\ref{fig:beta_temp_1}. However, \scite{mohr99} find that two-component
fits are required to adequately represent most cluster profiles, and that
the results from such fits show no trend in the value of $\beta_{fit}$ for
the extended cluster component. They conclude that results such as those of
\scite{arnaud99} arise from biases due to the inappropriate use of single
$\beta$-model profiles.  On the other hand, we {\it have} accounted for the
central component, but still find that $\beta_{fit}$ is substantially lower
in groups that the value of $2/3$ found for clusters by \scite{mohr99}.

The resolution of this situation probably lies in the temperature ranges
covered. The analysis of \scite{ponman99} is model-independent, in that it
involved simply overlaying the scaled surface brightness profiles.  This
shows that flattening of the profiles sets in at temperatures $T$\ltsim
3~keV. Since the sample of \scite{mohr99} includes only a single cluster
with $T<3$~keV, the lack of trend in $\beta_{fit}$ observed within their
sample, and the much flatter profiles observed in our sample, are both
consistent with the \scite{ponman99} results.

Finally, we wish to emphasize that an important implication of the flat
X-ray profiles of groups, coupled with their generally low surface
brightness compared to clusters, is that one must be very careful in
drawing conclusions about properties such as gas mass, gas fraction etc. on
the basis of analyses confined to `detection radii'. For example
\scite{mulchaey96} conclude that masses of gas in groups are typically
lower than the mass in galaxies, on the basis of analyses within the region
of detectable X-ray emission, which in many cases is only $\sim 200$~kpc.
Such results have important implications. For example, \scite{renzini97}
has used them to argue that the iron mass to light ratio in groups is much
lower than that in clusters, and that it is therefore difficult to explain
how clusters can be assembled through group mergers.

It can be seen from Fig.\ref{fig:fraction} that under the assumption that
our $\beta$-model fits can be extrapolated to $R_V$, less than 50\% of the
X-ray luminosity of the system is contained within 200~kpc for typical
groups. Now the asymptotic power law behaviour of surface brightness at
large $r$ is $S(r) \propto r^{1-6\beta}$, whilst the corresponding density
profile (in the approximation of isothermal gas) is $\rho_{gas}\propto
r^{-3\beta}$. Hence the density profile is even flatter, and the fraction
of the total gas mass contained within $r=$200~kpc will be considerably
{\it less} than 50\%.  The flat gas profiles mean that the gas fractions of
groups rise strongly with radius, so that very different results might be
obtained if our instruments were sufficiently sensitive to detect group
emission out to $R_V$, a possibility which should be realised with the
launch of XMM.

\section{Conclusions}
\label{Conclusions}

We have carried out detailed analysis of {\it ROSAT} PSPC data for 24 X-ray
bright galaxy groups. Temperatures and bolometric luminosities have been
derived for each group, and surface brightness profiles modelled in some
detail. In agreement with previous studies we find evidence for the
presence of two components in the surface brightness profiles of many of
the groups. When present, the central component is coincident with the
position of a central galaxy, suggesting that it may be due to the halo of
the galaxy, or to a cooling flow focused onto the central galaxy.

The surface brightness profiles of groups are significantly flatter than
those of galaxy clusters. For a subsample of the groups with the best data,
the steepness of the surface brightness profiles, as measured by the
parameter $\beta_{fit}$, appear to show a trend with mass when combined
with cluster data. This result is consistent with the idea that galaxy
winds have significantly affected the state of the intergalactic medium in
low mass systems.

The relation between the X-ray luminosity and temperature for galaxy groups
is also derived. This relation is found to be significantly steeper than
that derived for galaxy clusters. The action of galaxy winds flattening
surface brightness profiles would reduce the luminosity of the gas, due to
the luminosity dependance on the square of the density, thus causing a
steepening of the $L:T$ relation for lower mass systems. Further evidence
for this scenario is provided in the relation between velocity dispersion
and temperature. The $\sigma:T$ relation shows that for lower mass systems
the specific energy in the gas is greater than the specific energy in the
galaxies, suggesting that there has been energy injection in these systems.
An encouraging level of agreement is apparent between our results and
recent models and simulations of the effects of preheating by galaxy winds.

\section{Acknowledgements}
We thank Alex Deakin for his work in the early stages of this project,
John Mulchaey for interesting discussions about the X-ray properties of
groups, and the referee for suggesting several improvements to the paper.
Edward Lloyd-Davies and Bruce Fairley provided help and advice
on the data analysis and read several versions of this
manuscript.

SFH acknowledges financial support from the University of Birmingham. This
work made use of the Starlink facilities at Birmingham, the LEDAS database
at Leicester, the NASA/IPAC Extragalactic Database (NED), and images from
the STScI Digitized Sky Survey.

\bibliography{reffile}
\label{lastpage}

\end{document}